\documentclass[3p,times,procedia]{elsarticle}
\usepackage{nupha_ecrc}

\volume{00}

\firstpage{1}

\journalname{Nuclear Physics A}

\runauth{J. Ghiglieri}

\jid{nupha}

\jnltitlelogo{Nuclear Physics A}

\usepackage{amssymb}

\usepackage{graphicx}
\usepackage{amsmath}
\usepackage{amsfonts}
\usepackage{amssymb}
\usepackage{paralist}
\usepackage{slashed}
\usepackage{bm}
\usepackage{empheq}
\def\sss{\scriptscriptstyle}

\def \nc {N_c}
\def \ca {C_A}

\newcommand{\Tint}[1]{{\hbox{$\sum$}\!\!\!\!\!\!\!\int\,}_{\!\!\!\!\raise-0.9ex\hbox{$\scriptstyle{#1}$}}}
\def\onetwo{{1\lra2}}
\def\twotwo{{2\lra2}}

\def\siml{{\ \lower-1.2pt\vbox{\hbox{\rlap{$<$}\lower6pt\vbox{\hbox{$\sim$}}}}\ }}
\def\simg{{\ \lower-1.2pt\vbox{\hbox{\rlap{$>$}\lower6pt\vbox{\hbox{$\sim$}}}}\ }}

\def \bfnabla {\boldsymbol{\nabla}}

\def \als {\alpha_{\mathrm{s}}}

\def \m2   {\mu^{2 \epsilon}}

\def\siml{{\ \lower-1.2pt\vbox{\hbox{\rlap{$<$}\lower6pt\vbox{\hbox{$\sim$}}}}\ }}
\def\simg{{\ \lower-1.2pt\vbox{\hbox{\rlap{$>$}\lower6pt\vbox{\hbox{$\sim$}}}}\ }}

\def\qp {q_\perp}

\def\ql {\hat{q}_{\sss L}}
\def\etad{\eta_{\sss D}}

\def\lra{\leftrightarrow}

\def \OO {\mathcal{O}}

\def \qm {q^-}
\def \qll {q^+}

\def\x{{\bm x}}
\def\p{{\bm p}}

\def\v{{\bm v}}

\def\PP {f}

\def\qhat {\hat{q}}

\def\rrangle{\right\rangle}
\def\llangle{\left\langle}

\usepackage[figuresright]{rotating}

\begin{document}

\begin{frontmatter}



\dochead{}

\title{Energy loss at NLO in a
high-temperature Quark-Gluon Plasma
}


\author{Jacopo Ghiglieri}

\address{Institute for Theoretical Physics, Albert Einstein Center,
University of Bern, Sidlerstrasse 5, 3012 Bern, Switzerland}

\begin{abstract}
We present an extension of the Arnold-Moore-Yaffe kinetic equations 
for jet energy loss to NLO in the strong coupling constant. 
A novel aspect of the NLO analysis is a consistent description of wider-angle bremsstrahlung
(semi-collinear emissions), which smoothly interpolates between $\twotwo$ scattering
and collinear bremsstrahlung. 
We describe how many of the ingredients of the NLO transport equations 
(such as the drag coefficient) can be expressed in terms of Wilson line operators and
can be computed using a Euclidean formalism or sum
rules, both motivated by the analytic properties of amplitudes at light-like separations. 
We conclude with an outlook on the computation of the shear viscosity at NLO.

\end{abstract}


\end{frontmatter}


\section{Introduction}
\label{sec_intro}
Two main avenues for the investigation of the medium produced in heavy-ion collision 
are the study of its bulk properties on one hand and the analysis of hard probes on
the other. On the theory side, the former is mostly studied through an effective hydrodynamic
description, which kicks in after at some initial time $\tau_0\sim 1 fm/c$, after a rapid
\emph{thermalization} process has taken place. For what concerns hard probes,
considerable activity is dedicated to the investigation of \emph{jet quenching}. Theory
overviews of hydrodynamics,  thermalization and jet quenching formalisms
have been presented at this conference in \cite{gabriel}, \cite{aleksi,paul}, \cite{yacine}, 
with experimental reviews of flow and jet data in \cite{flowexp,jetexp}.
In this contribution we will concentrate on a weak-coupling theory approach that is well suited
to compute in-medium jet propagation, thermalization and the transport coefficients of the QCD
medium. It is the \emph{effective kinetic theory} derived by Arnold, Moore and Yaffe (AMY) 
\cite{Arnold:2002zm} and used for the leading-order computation of the transport coefficients,
such as the shear viscosity, in \cite{Arnold:2003zc}. In this contribution
we will show how the version of this kinetic theory suited to the study
of jet propagation can be extended to the next order in the strong coupling 
$g=\sqrt{4\pi\als}$, effectively giving a summary of the results presented in
detail in \cite{Ghiglieri:2015ala} and introduced more pedagogically 
in \cite{Ghiglieri:2015zma}. One important motivation for this extension to
NLO is to gauge the stability of perturbation theory, which requires $g\ll 1$ 
at finite temperatures, when extrapolated to temperatures where $\als\sim 0.3$.

This contribution is organized as follows: in Sec.~\ref{sec_coll}
the LO kinetic theory is reviewed and a useful reorganization of
its collision operator is introduced and extended to NLO in the case of jet quenching,
while Sec.~\ref{sec_eta} contains an outlook on the extension to 
transport coefficients and a brief conclusion.

\section{Reorganization of the collision operator}
\label{sec_coll}
The AMY kinetic theory can be written as
\begin{equation}
	\left(\partial_t+\v\cdot\bfnabla_\x\right)f(\p)=C^\twotwo+C^\onetwo,
	\label{amy}
\end{equation}
where the l.h.s. is the typical one for a Boltzmann equation in the absence of 
external forces,while the r.h.s. is the \emph{collision operator}, written
as a sum of $\twotwo$ and $\onetwo$ components. The former are the standard
elastic scatterings of a gauge theory, complemented by Hard Thermal Loop (HTL)
resummation \cite{Braaten:1989mz} for IR finiteness. $\onetwo$ labels
$1+n\lra 2+n$ processes, i.e. the collinear splittings/joinings of one particle into
two other, induced by $n\ge 1$ soft scatterings with medium constituents. The 
coherent, destructive interference of these scatterings gives rise to the well-known 
Landau-Pomeranchuk-Migdal (LPM) effect.
For ease of illustration, we will omit quarks entirely
in this contribution, i.e. we consider energetic gluons propagating through an equilibrated
gluon plasma at a temperature $T$. $f(\p)$ denotes
the phase space distribution of the jet gluons and the collision operator can be linearized
in $f$, i.e. only one of the 3 or 4 particles in a $\onetwo$ or $\twotwo$ process
is a jet parton, while all others are thermal and characterized by the equilibrium distribution
Bose-Einstein $n_\mathrm{B}$.

\begin{figure}[ht]
	\vspace{-0.3cm}
	\begin{center}
		\includegraphics[width=3cm]{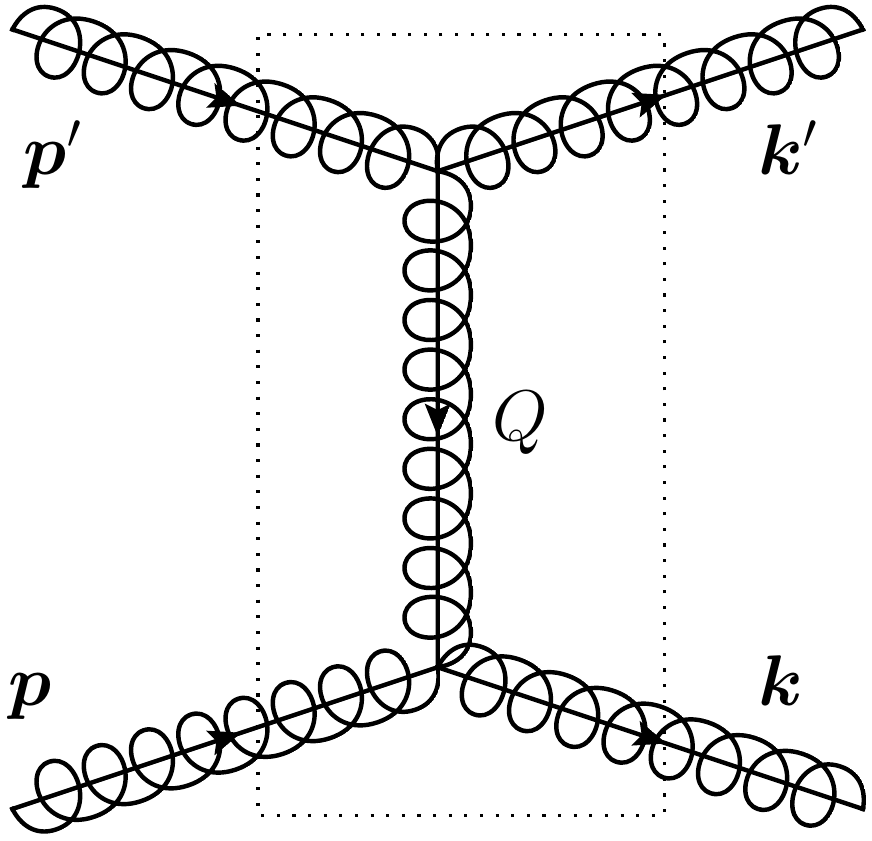}
		\includegraphics[width=7cm]{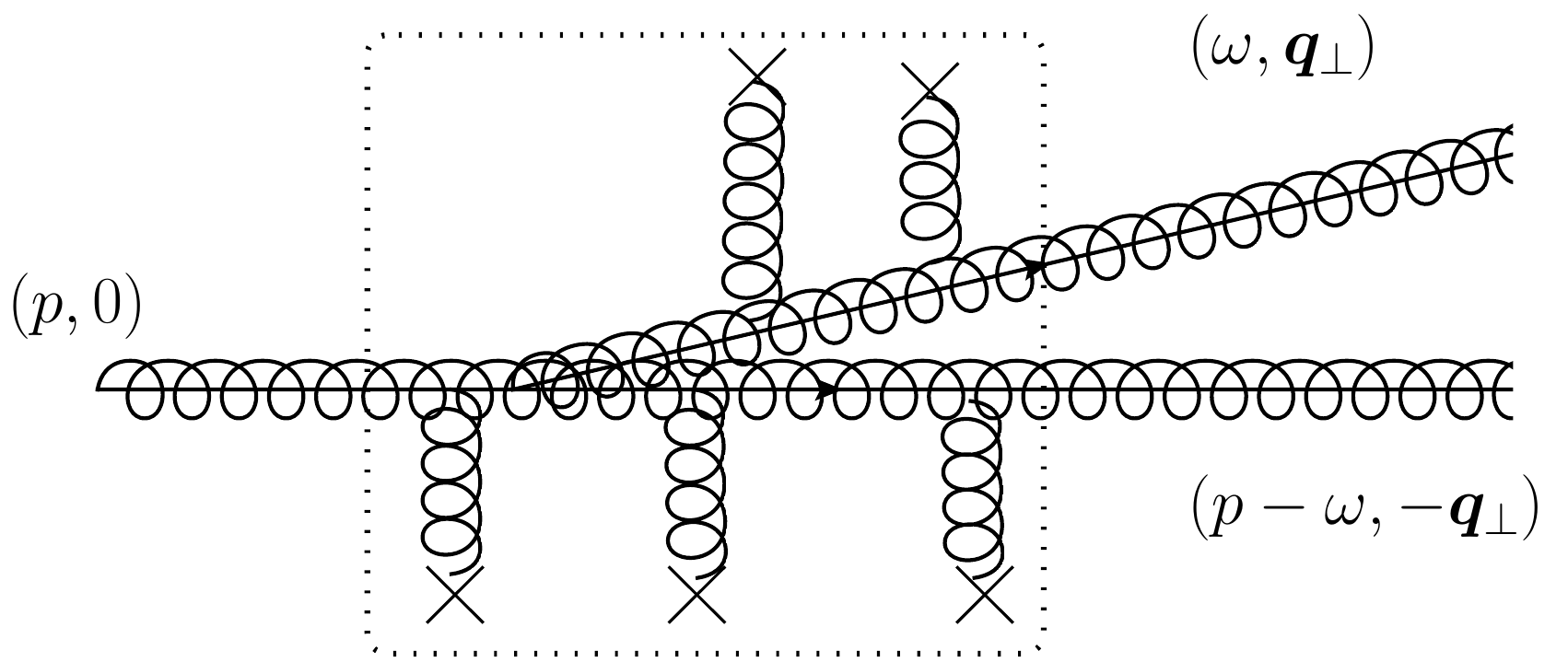}
	\end{center}
	\vspace{-0.5cm}
	\caption{Left: a large-angle scattering process. Right:
		a collinear process. In both cases, curly
	lines with superimposed straight, solid line represent
gluons whose energy and momentum are hard, i.e. at least of order $T$. 
Curly lines	without superimposed straight lines are soft gluons, 
		and the crosses represent the thermal gluons they
	scatter from.The boxed
area represent the process entering the kinetic equation.}
\label{fig_large}
\end{figure}

\begin{figure}[ht]
	\vspace{-0.3cm}
	\begin{center}
		\includegraphics[width=2cm]{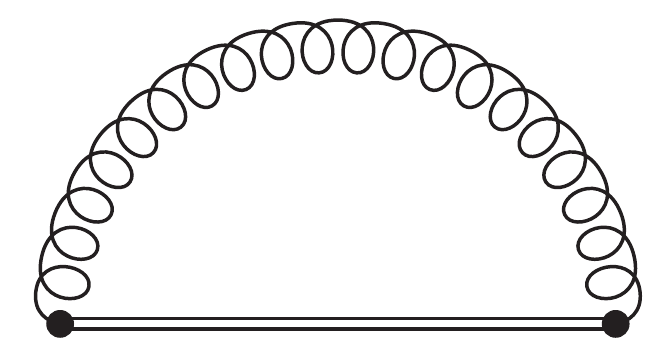}
		\includegraphics[width=10cm]{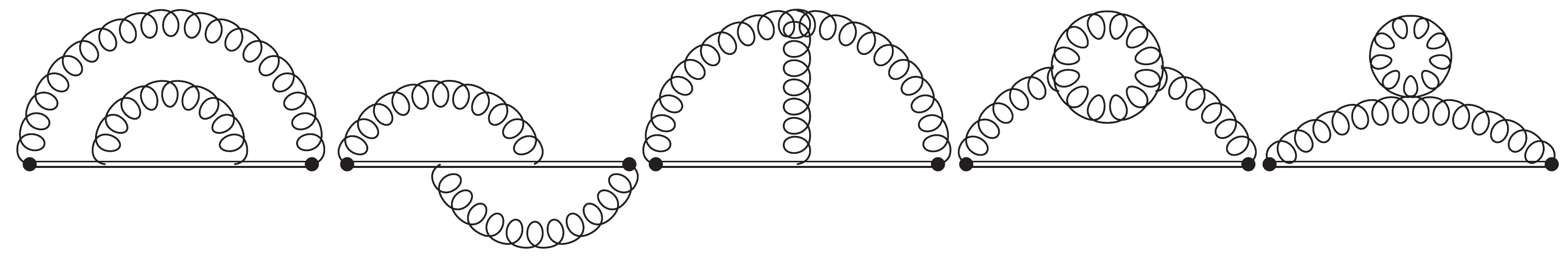}
	\end{center}
	\vspace{-0.5cm}
	\caption{Diagrams for the evaluation of the longitudinal
	and transverse momentum diffusion coefficients at LO (first one)
and NLO (others). The dots represent the field strength insertions, the
double line is the adjoint Wilson line connecting them and curly lines
are HTL soft gluons.}
	\label{fig_diffusion}
\end{figure}

The separation between $\onetwo$ and $\twotwo$ processes ceases to be well defined beyond
leading order. In order to have a collision operator that is more easily extended to NLO and
which makes more transparent the introduction of effective Wilson line descriptions for soft
scattering processes, we introduce the following reorganization
\begin{equation}
	\left(\partial_t+\v\cdot\bfnabla_\x\right)f(\p)=C^{\mathrm{large}}[\mu_\perp]
	+C^\mathrm{diff}[\mu_\perp]+C^\mathrm{coll},
	\label{reorganize}
\end{equation}
where the three processes are \emph{large-angle processes}, \emph{diffusion processes}
and \emph{collinear processes}. A large-angle process is a $\twotwo$ process with an 
$\OO(1)$ angular deflection, which translates into a large momentum transfer $Q\simg T$,
enforced by an infrared cutoff $\mu_\perp$.
An example is depicted on the left in Fig.~\ref{fig_large}. A collinear process is instead
a $\onetwo$ process with strictly collinear kinematics: at leading order it coincides
with $\onetwo$ processes, but at NLO a subtraction of the limits where it blurs with other
processes is required. It is represented on the right in Fig.~\ref{fig_large}.
Finally, the region of soft momentum exchanges in $\twotwo$ processes is described
in a diffusion picture, i.e.
\begin{equation}
	\label{diff}
C^\mathrm{diff}[\mu_\perp]\equiv-\frac{\partial}{\partial p^i}\bigg[\etad(p)p^i \PP(\p)\bigg]
-\frac12\frac{\partial^2}{\partial p^i\partial p^j}
\left[\left(\hat{p}^i\hat{p}^j\ql+\frac12(\delta^{ij}-\hat{p}^i\hat{p}^j)\qhat
\right)\PP(\p)\right],
\end{equation}
where the three coefficients entering in this Fokker-Planck equation
are the drag $\eta_D$ and the longitudinal and transverse momentum broadening coefficients
$\ql$ and $\qhat$. These two can be defined as effective force-force correlators
on Wilson lines along the classical trajectories of the particles, i.e.
\begin{equation}
	\label{defq}
\hat q^{ij}  \equiv  \int_{-\infty}^{+\infty} dt ' \,\llangle\mathcal F^i(t) \mathcal F^j(0)\rrangle,\,
	\qquad
	\mathcal F^{i}(x^+)  \equiv  U^\dagger(x^+,-\infty) \, g F^{i\mu}(x^+)v_{\mu} \, U(x^+,-\infty) \, .
\end{equation}
where $U$ is an adjoint Wilson line in the $x^+\equiv(x^0+x^z)/2$ light-cone direction ($x^-\equiv x^0-x^z)$
in which we have taken the energetic jet particle to be. Similarly, $v^\mu=(1,0,0,1)$ is a null
vector pointing in that same direction.
At leading order this operator takes the form depicted in the first diagram on the left in 
Fig.~\ref{fig_diffusion}. The LO $\qhat$ can be easily evaluated using the mapping to
the three-dimensional Euclidean theory introduced by Caron-Huot \cite{CaronHuot:2008ni}
and reviewed in \cite{Ghiglieri:2015zma}, yielding
\begin{equation}
	\qhat=g^2\ca\int^{\mu_\perp}\frac{d^2\qp}{(2\pi)^2}\int\frac{d\qll}{2\pi}
	\llangle F^{-\perp}(Q)F^-_{\;\perp}\rrangle_{\qm=0}
	=g^2\ca T\int^{\mu_\perp}\frac{d^2\qp}{(2\pi)^2}\frac{m_D^2}{\qp^2+m_D^2}
	=\frac{g^2\ca Tm_D^2}{2\pi}\ln\frac{\mu_\perp}{m_D},
	\label{qhat}
\end{equation}
where $m_D^2=\nc g^2T^2/3$ is the Debye mass.
The longitudinal one can instead be evaluated using a new sum rule,
based on the analytical properties of amplitudes at space- and light-like separations
\cite{Ghiglieri:2015ala}, yielding
\begin{equation}
	\ql=g^2\ca\int^{\mu_\perp}\frac{d^2\qp}{(2\pi)^2}\int\frac{d\qll}{2\pi}
	\llangle F^{-z}(Q)F^-_{\;z}\rrangle_{\qm=0}
	=g^2\ca T\int^{\mu_\perp}\frac{d^2\qp}{(2\pi)^2}\frac{m_\infty^2}{\qp^2+m_\infty^2}
	=\frac{g^2\ca Tm_\infty^2}{2\pi}\ln\frac{\mu_\perp}{m_\infty},
	\label{qhatl}
\end{equation}
where $m_\infty^2=m_D^2/2$ is the asymptotic mass  of gluons.
Finally, $\eta_D$ can be determined from the other two through an Einstein-like
relation, obtained by imposing that the Fokker-Planck picture be equivalent
to the Boltzmann one for $Q\gg gT$ and that it show a fixed point at equilibrium. The
UV logarithmic dependence of the diffusion sector cancels with the IR one
in large-angle scattering processes.

When considering higher-order terms, 
soft gluon loops, thanks to the Bose enhancement $n_\mathrm{B}(gT)\sim 1/g$, are only
suppressed by $g$, rather than $g^2$. This implies that the collinear and diffusion sector
receive $\OO(g)$ corrections from the inclusion of soft gluon loops. Furthermore, a new,
\emph{semi-collinear process} has to be considered at NLO.

In the collinear sector, the soft scattering rate $d\Gamma/(d^2\qp)$ inducing the splitting
process, receives $\OO(g)$ corrections from these loops. The Euclidean mapping mentioned before
was indeed first applied to the computation of this correction \cite{CaronHuot:2008ni}.
A similar soft correction to the dispersion relation \cite{CaronHuot:2008uw}
of the collinear particles needs also to be considered. 

For diffusion, the two-loop diagrams depicted in Fig.~\ref{fig_diffusion} need to be computed
to obtain the $\OO(g)$ corrections to $\qhat$ and $\ql$. The former (related to $d\Gamma/(d^2\qp)$)
was done in \cite{CaronHuot:2008ni} using the Euclidean mapping, while the latter \cite{Ghiglieri:2015ala}
simplifies greatly using the aforementioned sum rule, resulting in the replacement of $m_\infty^2$ in 
Eq.~\eqref{qhatl} with $m_\infty^2+\delta m_\infty^2$, where $\delta m_\infty^2$ is the $\OO(g)$
correction to the asymptotic mass. 

\begin{figure}[ht]
	\begin{center}
		\includegraphics[width=10cm]{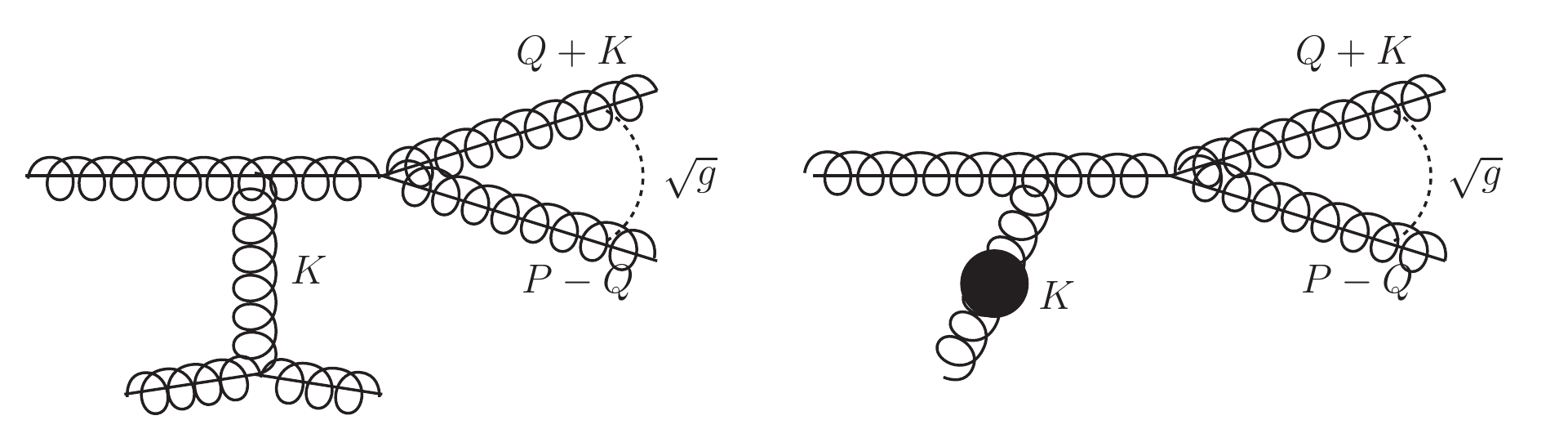}
	\end{center}
	\vspace{-0.5cm}
	\caption{Semi-collinear processes. On the left the splitting is induced
	by a soft scattering, while on the right by the absorption of a plasmon 
(the black blob).}
	\label{fig_semi}
\end{figure}

Finally, semi-collinear processes can be seen as collinear processes with larger opening
angles. This reduces the collinear enhancement, making them subleading. Furthermore, the relaxed
kinematical constraints also allow the interactions with plasmons, besides the usual space-like
soft scatterings, as shown in Fig.~\ref{fig_semi}.
 The evaluation of these diagrams proceeds similarly to that of the
collinear ones in the single-scattering (Bethe-Heitler) limit, where LPM interference
is suppressed. The rate turns out to be proportional to the DGLAP splitting kernel
multiplying a generalized $\qhat$ which keeps track of the component
of the soft gluon momentum in the other light-cone direction. We call it $\qhat(\delta E)$
and it reads
\begin{equation}
	\qhat(\delta_E)=g^2\ca\int^{\mu_\perp}\frac{d^2\qp}{(2\pi)^2}\int\frac{d\qll}{2\pi}
	\llangle F^{-\perp}(Q)F^-_{\;\perp}\rrangle_{\qm=\delta E}.
	\label{qhatde}
\end{equation}
It too can be evaluated using the Euclidean mapping. An IR log divergence
in these processes cancels exactly an UV one in diffusion processes.

In order to establish the quantitative effect of the NLO corrections
we have just obtained, a numerical implementation in a Monte Carlo
generator such as MARTINI \cite{Schenke:2009gb}, which currently implements $\onetwo$ and
$\twotwo$ processes at LO, is underway.
\section{Outlook on transport coefficients and conclusions}
\label{sec_eta}
\begin{figure}[ht]
	\vspace{-0.3cm}
	\begin{center}
		\includegraphics[width=5cm]{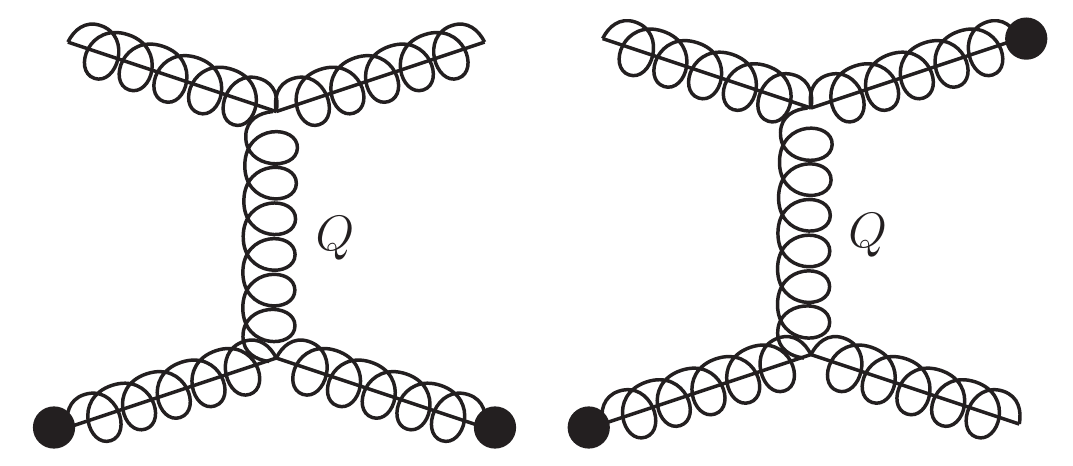}
	\end{center}
	\vspace{-0.5cm}
	\caption{The dots represent the $T^{ij}$ insertions.}
	\label{fig_eta}
\end{figure}

The techniques we have briefly illustrated,
which allow to cast all the intricate soft dynamics into a few effective 
operators evaluated using Euclidean mappings or sum rule, can in
principle be applied to transport coefficients as well. There is however
one extra major difficulty in that case. Consider the computation
of the shear viscosity $\eta$: it requires knowing how a disturbance in 
the energy-momentum tensor $T^{ij}$ sources a second $T^{ij}$ disturbance. In
other words, one needs a linearized kinetic theory where two off-equilibrium
distributions are considered within the same process, rather than one. In the case
of a soft $\twotwo$ scattering, one has then the two possibilities in Fig.~\ref{fig_eta}. 
In the leftmost case, the two $T^{ij}$ insertions are on the same side of the soft gluon. 
Their momenta are thus strongly correlated, as they differ by $Q\sim gT$ and a diffusion picture
similar to the one we have introduced is applicable. In the rightmost case, on the other hand,
the momenta are uncorrelated and the diffusion picture is not applicable. An inspection
of the leading-order calculation shows however that such terms are UV finite, making the
prospect of a direct, ``brute-force'' NLO calculation in the HTL theory slightly less daunting.

In conclusion, the reorganization of the kinetic theory, which recasts all soft contributions in
light-front Wilson-line operators, is an extremely useful tool, which is now generalized to NLO
for jet propagation. The extension to transport coefficient is more problematic, but a reliable 
estimate should be feasible with the current technology.\\
\emph{Acknowledgements} I thank G.~Moore and D.~Teaney for collaboration.
	My work is supported by the Swiss
National Science Foundation (SNF) under grant 200020\_155935.





\bibliographystyle{elsarticle-num}
\bibliography{eloss.bib}







\end{document}